\documentclass[journal]{IEEEtran}

\usepackage{xcolor,soul,framed} 

\colorlet{shadecolor}{yellow}
\usepackage[pdftex]{graphicx}
\graphicspath{{../pdf/}{../jpeg/}}
\DeclareGraphicsExtensions{.pdf,.jpeg,.png}

\usepackage[cmex10]{amsmath}
\usepackage{array}
\usepackage{mdwmath}
\usepackage{mdwtab}
\usepackage{eqparbox}
\usepackage{url}
\usepackage{stfloats}
\usepackage{cite}
\usepackage{tabularx}
\begin{document}
    \title{Molecular Communication-Based Intelligent Dopamine Rate Modulator for Parkinson's Disease Treatment
}
  \author{Elham~Baradari,~\IEEEmembership{Student Member,~IEEE,}
      ~Ozgur~B~Akan,~\IEEEmembership{Fellow,~IEEE}

  \thanks{Elham Baradari is with the Center for NeXt-Generation Communications (CXC), Department of Biomedical Sciences and Engineering, Koç University, 34450 Istanbul, Turkey  (e-mail: {ebaradari22, akan}@ku.edu.tr).}
  \thanks{Ozgur B. Akan is with the Internet of Everything (IoE) Group, Department of Engineering, University of Cambridge, CB3 0FA Cambridge, U.K., and also with the Center for NeXt-Generation Communications (CXC), Koç University, 34450 Istanbul, Turkey (e-mail: oba21@cam.ac.uk).}}%

\maketitle

\begin{abstract}
Parkinson’s disease (PD) is a progressive neurodegenerative disease, and it is caused by the loss of dopaminergic neurons in the basal ganglia (BG). Currently, there is no definite cure for PD, and available treatments mainly aim to alleviate its symptoms. Due to impaired neurotransmitter-based information transmission in PD, molecular communication-based approaches can be employed as potential solutions to address this issue. Molecular Communications (MC) is a bio-inspired communication method utilizing molecules for carrying information. This mode of communication stands out for developing bio-compatible nanomachines for diagnosing and treating, particularly in addressing neurodegenerative diseases like PD, due to its compatibility with biological systems. This study presents a novel treatment method that introduces an Intelligent Dopamine Rate Modulator (IDRM), which is located in the synaptic gap between the substantia nigra pars compacta (SNc) and striatum to compensate for insufficiency dopamine release in BG caused by PD. For storing dopamine in the IDRM, dopamine compound (DAC) is swallowed and crossed through the digestive system, blood circulatory system, blood-brain barrier (BBB), and brain extracellular matrix uptakes with IDRMs. Here, the DAC concentration is calculated in these regions, revealing that the required exogenous dopamine consistently reaches IDRM. Therefore, the perpetual dopamine insufficiency in BG associated with PD can be compensated. This method reduces drug side effects because dopamine is not released in other brain regions. Unlike other treatments, this approach targets the root cause of PD rather than just reducing symptoms.
\end{abstract}



%
\IEEEpeerreviewmaketitle


\section{Introduction}

\IEEEPARstart{P}{ARKINSON's} disease (PD) is a progressive neurodegenerative disorder that is prevalent among individuals over the age of 60 \cite{olanow2009scientific,przedborski2017two}. Neurodegenerative disorders are a categorization of neurological diseases characterized by the formation of distinct brain lesions that progressively degenerate over time. The presence of brain lesions, together with the progressive decline in neurocentral control, is the underlying cause of the advancing symptoms observed in patients. Movement disorders, psychological and cognitive impairments generally characterize PD \cite{chaudhuri2009non,bohnen2022discussion, aarsland2021parkinson}. The disease is caused by the death of dopaminergic neurons in the substantia nigra pars compacta (SNc) region of the basal ganglia (BG). PD currently has no definitive cure, and the current treatments, such as dopaminergic drugs\cite{tran2018levodopa}, and deep brain stimulation (DBS) \cite{dale2022evoked, haeri2005modeling}, can only alleviate its symptoms.

Several studies have addressed PD, with some aiming to identify the symptoms of the disease for early diagnosis \cite{armstrong2020diagnosis, little2021smartphones}, and others striving to develop conceptual and quantitative models \cite{di2018quantitative}. However, most research in this field has attempted to propose alternative treatment approaches that do not involve pharmaceuticals or DBS \cite{bloem2015nonpharmacological, wang2022efficacy}. Some studies focused on nano-scale structures for new neurodegenerative disease treatments \cite{modi2009nanotechnological, spuch2012advances}. Enabling communication between nano-scale devices may significantly increase their capabilities for treatment purposes \cite{biletic2020nanonetworks}. Molecular communication (MC), which involves the encoding, transmitting, and receiving of information using molecules, enables the development of bio-compatible nanomachines for developing novel diagnosis and treatment strategies for the dysfunctions of intra-body nanonetworks \cite{akan2016fundamentals, dressler2010bio}. This is regarded as the most favorable method of communication among bioinspired nanomachines. Furthermore, the biocompatible MC-based nanonetworks make them applicable to various biological applications \cite{atakan2012body, atakan2012bio}. One of the most promising nano-scale communication paradigms is communication among neurons \cite{vegh2022towards}. The utilization of nanomachines for treating neurodegenerative diseases, such as PD, enables the interface between neurons and nanomachines. However, therapies for these diseases still face significant challenges, highlighted in recent studies \cite{veletic2019molecular}, and innovative approaches based on molecular engineering and nanotechnology are being developed to address these issues. Neurological diseases have been thoroughly studied from the perspective of MC \cite{akan2021information}. Thus, it appears that MC-based techniques may be beneficial for PD. Although the primary root cause of PD is the loss of SNc neurons, and there is no definitive therapy, MC strategies offer promise in providing a potential solution. These methods can be particularly beneficial for advanced stages of PD, where drug therapy becomes less effective due to drug resistance. While invasive DBS-based treatments have also proven effective in some cases, they only provide symptom relief, highlighting the need for a more fundamental treatment based on MC.

 This study aims to introduce a novel MC-based method for treating PD. In both therapy approaches (drug and DBS), attention is not directed toward the underlying cause of the disease, which is defined by the degeneration of dopaminergic neurons in the SNc. Achieving a complete treatment for the disease requires substituting dying neurons for newly generated ones, a pursuit that is currently unachievable. On the other hand, the potential replacement of dead neurons with artificial neurons is currently being considered. Understanding the communication theoretical capabilities of information transmission among neurons, known as neuro-spike communication, is a significant step in developing bio-inspired solutions for nano networking \cite{dressler2010survey, malak2012molecular}. Communication between neurons occurs via transmission of neural spike trains through junctional structures, either electrical or chemical synapses, providing connections among nerve terminals \cite{malak2013communication, ramezani2017communication, ramezani2018impacts, ramezani2017importance, khan2019impact}. However, simulating the exact function of biological neurons in constructing an artificial neuron currently has significant obstacles. If it were feasible to replace the critical functionalities of neurons using a system, such as nanomachines, this suggests a promising approach for more effective disease management. Considering the crucial function of a neuron in receiving stimuli from presynaptic neurons and the targeted release of neurotransmitters to influence the subsequent postsynaptic neurons, it is reasonable to propose nanomachines capable of detecting the activation of SNc neurons and addressing dopamine reduction caused by the loss of neurons in this area. In light of these particular circumstances, a notable achievement will be attained in pursuing a fundamental treatment for PD. 
 
 This paper is organized as follows. In Section II, the physiological background and the problems caused by PD, as well as its current treatments are identified. Following this, Section III attempts to propose ideas for solving the challenges of PD to compensate for dopamine deficiency, and the foundational principles underlying an intelligent dopamine rate modulator (IDRM) design for providing exogenous dopamine in BG are explained. Section IV aims to explain the mechanism of DDS in order to charge dopamine in IDRMs. The drug concentration in each region has been meticulously calculated and estimated considering the dynamics of substance absorption in the digestive system, blood circulation, passage through the blood-brain barrier, and intracerebral distribution. The results indicate that with the intake of dopamine-containing tablets, the required amount of dopamine can be supplied to the ‌‌BG. Moving forward, Section V presents numerical results accompanied by their corresponding results. Finally, Section VI describes the general summary of the study.


\section{Physiological Background}

\begin{figure}
  \centering
  \includegraphics[width=\columnwidth]{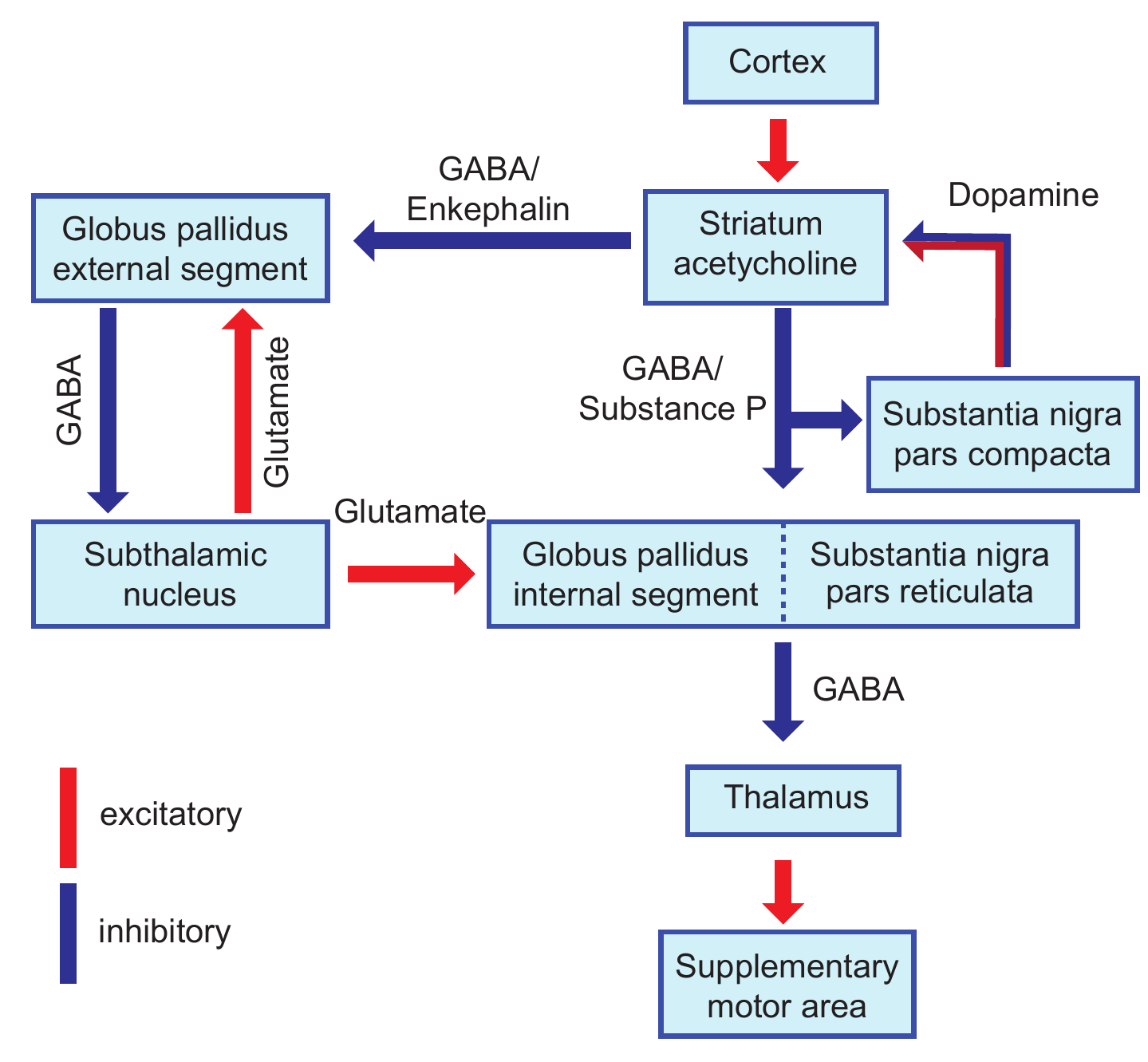}
  \caption{Basal Ganglia (BG) Pathways: Direct and Indirect Routes, along with Neuronal Connections and Neurotransmitters \cite{haeri2005modeling}.}\label{BG_model}
\end{figure}

The basal ganglia (BG) is a crucial area in the brain that plays a critical role in both motor and non-motor activities. Dysfunction in the basal ganglia can lead to various neurological disorders, such as Parkinson's disease, Huntington's disease, and hemiballismus, each with distinct consequences. 
The BG consists of multiple neuronal blocks that intricately interact, forming a complex network of interconnected neural structures that mutually influence one another. Fig.~\ref{BG_model} depicts the constituent elements of the BG, the interplay between neuronal units, and the specific neurotransmitter connections involved.

Following the conceptual model illustrated in Fig.~\ref{BG_model}, the BG receives input from the cortex and subsequently transmits efferent outputs back to the cortex via the thalamus. Information processing in the BG can occur through two distinct paths: the direct and indirect pathways. Upon detecting a reward signal, a limited amount of processing is required, activating the direct path. On the contrary, the indirect method is selected when there is a need for more processing. The mechanism for selecting the direct and indirect pathways is regulated by the dopaminergic signals originating from the substantia nigra pars compacta (SNc), which are responsible for processing reward-related stimuli. This function serves as an implementation of reinforcement learning, constituting the fundamental mechanism of the operation of the BG. PD is characterized by the loss of neurons in the SNc, resulting in a significant disruption of dopaminergic reward transmission \cite{lang2004challenges}. In recent years, drug-based therapies for PD have increased dopamine levels not only in the BG but also in the entire brain. Although this compensates for dopamine deficiency in the basal ganglia, it disrupts the functioning of other dopaminergic regions in the brain. Therefore, despite reducing PD symptoms, it leads to long-term side effects. Indeed, it is accurate to assert that in the early stages, medication suppresses the symptoms of the disease.
Nevertheless, the existence of exogenous dopamine in the neural pathway linking the SNc and the striatum suggests a persistent reinforcement signal that interferes with the proper operation of the BG. The reward signal carries crucial information that cannot be eliminated. The significance of the dopamine transporter in transmitting this information is apparent. To preserve the BG's function in PD, applying the principles of the molecular communication system for therapy seems necessary. Through this approach, vital dopaminergic information can be delivered to the striatum. DBS therapy involves the targeted delivery of electrical stimulation to precise regions inside the brain, thereby interfering with the pathological manifestations associated with the disease within the BG. However, despite its ability to control symptoms, it disrupts the normal functioning of the BG once more.

\begin{figure}
  \begin{center}
  \includegraphics[width=\columnwidth]{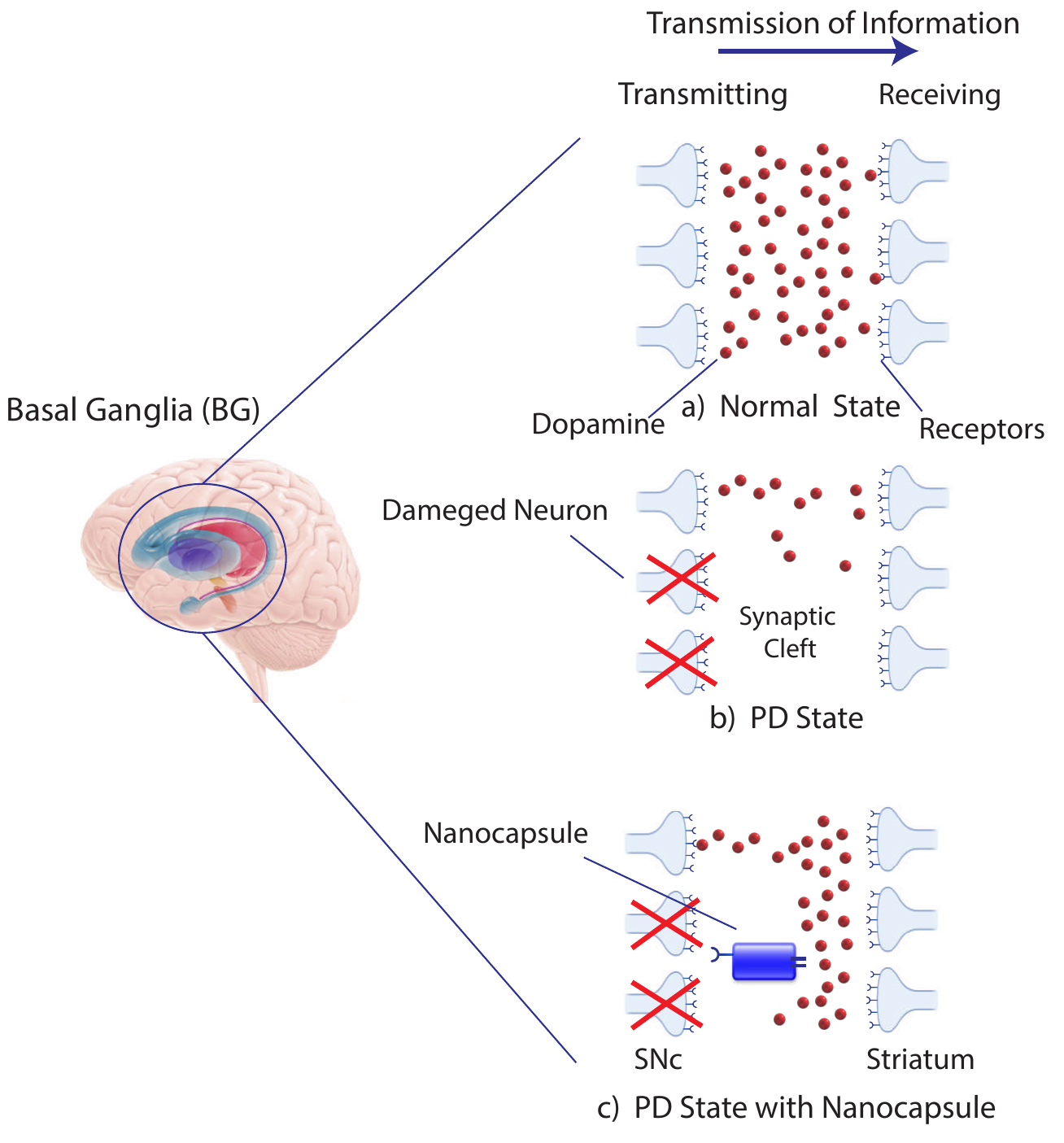}\\
  \caption{Dopamine levels in three different states: a) Healthy state, b) PD state, and c) Hypothesis regarding IDRMs for alleviating the PD state.}\label{3state}
  \end{center}
\end{figure}

By developing an IDRM that can identify dopamine release from the SNc and address dopamine insufficiency caused by PD, it becomes feasible to effectively maintain the normal functioning of the BG and mitigate the symptoms associated with PD. Fig.~\ref{3state} depicts the primary mechanisms underlying this hypothesis. Fig.~\ref{3state}(a) illustrates a synaptic connection between the SNc and the striatum in a state of health. This scenario involves the transmission of sufficient dopamine to the striatum upon stimulation of the SNc. Fig.~\ref{3state}(b) depicts the state of PD. The disruption of the dopamine reward signal occurs due to the loss of a number of SNc neurons, leading to inadequate signal generation despite SNc stimulation. The status of a patient with PD who has undergone treatment with IDRMs is depicted in Fig.~\ref{3state}(c).
In contrast to instances involving neuronal degeneration in the SNc, including IDRMs facilitates the appropriate generation of dopaminergic signals. In the given situation, the functionality of the BG remains primarily intact, resulting in the patient's behavior closely resembling that of people without any impairments. Following that, Section III investigates the feasibility and implementation of the proposed method.

\section {designing Intelligent
Dopamine Rate Modulator for Parkinson’s disease treatment}

Here, we introduce a technique that provides dopamine when needed in the BG region to approximate the behavior of individuals with PD to that of healthy individuals. Including a detection mechanism for dopamine release and integrating a storage mechanism for dopamine inside IDRM are crucial aspects to consider in designing this component. For this reason, IDRM must encompass two discrete components: A storage unit and an intelligent mechanism for dopamine release, as shown in Fig. \ref{receivers}. The proposed IDRM is located in the synaptic gap connecting the SNc and the striatum. Its primary objective is to effectively restore the functional impact of SNc dopaminergic neurons on the striatum. Following this, the design process responsible for the intelligent release of dopamine will be clarified, and subsequently, the mechanism involved in storing dopamine will be explained.

\subsection{Dopamine Release System}

The primary purpose of the IDRM is to provide the controlled and timely release of dopamine as required. An attempt will be undertaken to utilize biological activities to establish the appropriate timing for release. Based on biological findings, it has been observed that the SNc neurons become depolarized when dopamine is required, producing action potentials that propagate along their axons. Consequently, this process initiates the release of dopamine vesicles, leading to the subsequent diffusion of dopamine into the synaptic cleft. This segment effectively embodies biological MC systems, where the SNc neurons serve as the transmitters, the striatum neurons act as the receivers, and the synaptic cleft functions as the channel. From a practical standpoint, dopamine molecules function as carriers of information. It has been demonstrated that PD is characterized by deficient transmitter components, specifically the SNc neurons, resulting in insufficient dopamine synthesis.
Consequently, the IDRMs must function as a mechanism to address dopamine deficiency. Nevertheless, the restoration of dopamine must be carried out in a manner that effectively conveys relevant information. Accordingly, the release of dopamine should occur exclusively in response to a specific need. Typically, dopamine release is closely associated with the activation of the SNc. 

In the context of PD, there is a notable reduction in the number of neurons in the SNc. However, a subset of neurons manages to persist. The innate ability of these functioning neurons to detect stimuli and subsequently release dopamine is readily apparent.
Nevertheless, the demise of neurons will precipitate a decline in dopamine synthesis, resulting in a state of insufficiency. The intended should demonstrate the ability to detect the physiological secretion of dopamine. This implies that when there is stimulation inside the SNc, and there is a need to release an appropriate amount of dopamine, the IDRM, upon recognizing the presence of endogenous dopamine, proceeds to release a portion of its stored dopamine. Under these circumstances, the receptors within the striatum receive an adequate quantity of dopamine, ensuring the proper conveyance of information. The mechanism is depicted in Fig.~\ref{3state}. As can be inferred, using IDRMs emulates the receptor behavior within the striatum in PD, closely resembling that in healthy individuals.

\subsection{Dopamine Sensor}

Since dopamine is a vital neurotransmitter in the body, biologically, dopamine receptors are present in postsynaptic neurons. Receptors can be conceptualized as protein channels that exhibit sensitivity to ligands, such as neurotransmitters, causing the ion channel gate to open upon ligand binding. When exposed to dopamine, these receptors undergo conformational changes leading to channel opening, thereby enabling the selective passage of ions, such as sodium ions, which are abundant in the extracellular matrix. Consequently, a dopamine sensor can be envisioned as a biologically inspired simulation protein akin to dopamine receptors \cite{xie2019nanomaterial}, which, upon activation, permits the translocation of a specified amount of sodium ion.

\subsection{Intelligent Dopamine-releasing}

The attempt to develop an intelligent dopamine release mechanism for the IDRM aligns with creating highly selective biological channels responsive to specific ions \cite{lu2023artificial}. One end of these channels is located within the dopamine storage compartment, while the other is positioned outside the IDRM. As a result, in the presence of sodium ions, the channel will experience a specific time of activation, which enables the release of stored dopamine. The IDRM's output will be near the dopamine receptors within the striatum, releasing the necessary dopamine in situations requiring dopamine transmission.

\subsection{Dopamine Storage}

The previous sections have explained the method used to determine the requirement for dopamine and the process by which it is released from the IDRM. One of the critical challenges is the process of delivering and maintaining dopamine within the IDRM. To mitigate any disruption of endogenous dopamine in other cerebral regions, it is imperative to employ a dopamine molecule during the charging procedure of the IDRM. This compound should undergo conversion to dopamine exclusively upon entering the capsule. 




The ultimate goal is administering dopamine to the SNc and the striatum region. 
The reception of this new compound necessitates the presence of IDRM receptors, and their structural design will be contingent upon the specific dopamine compound (DAC) of interest \cite{alabrahim2022polymeric}. Following the ingress of the DAC into the capsule's interior, a designated filtration zone is strategically placed within the pathway of the DAC. The primary function of this filtration zone is to separate and convert any additional branches into dopamine. The produced DAC are enclosed within vesicles to facilitate their storage. Subsequently, in cases where dopamine release requires dopamine, the dopamine contained within these vesicles will undergo exocytosis and be released into the synaptic space next to the striatum. Fig. \ref{receivers} provides a schematic representation elucidating the capsule's release and charging segments.

It is imperative to clarify that transporting substances (specifically dopamine) within the releaser segment is inherently connected to transmitting physiological information. Cerebral internal mechanisms intricately regulate this physiological information. Conversely, transporting substances (in the form of DAC) predominantly assumes a nutritive role and do not inherently encompass discernible informational content. Thus, the releaser component can be aptly perceived as an integral element of the biological MC system, instrumental in participating in the body's natural processes with a distinct focus on information propagation. Conversely, despite utilizing MC mechanisms, the charger segment does not merit inclusion within the MC system framework owing to its lack of information transmission capabilities.

\section{IDRM Receiver Types: The Passive and Absorbing Receivers}

IDRM receivers play an essential role in the functionality of nanomachines \cite{kilinc2013receiver}, thereby requiring the exploration of many receptor types. The IDRM has two distinct types of receivers. The initial component is a sensory receiver situated within the releaser region, responsible for detecting the existence of endogenous dopamine. The second receiver is positioned within the charging compartment, transmitting DAC and subsequently delivering it into the capsule. Due to these different functions, these two receivers will have distinct structures \cite{kuscu2019transmitter}. 

\begin{figure*}[t]
  \centering
  \includegraphics[width=\textwidth]{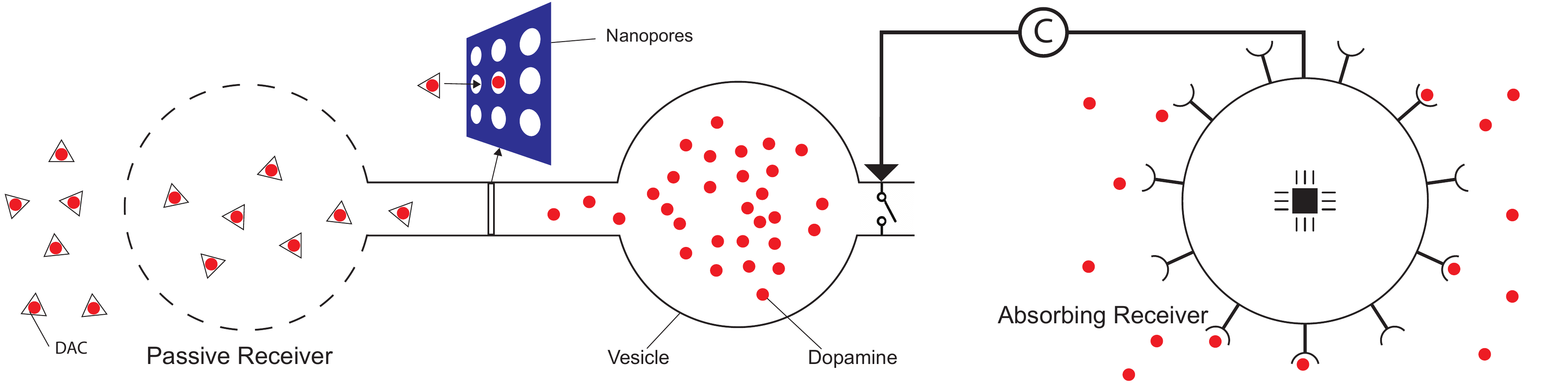}
  \caption{Proposed Receivers for IDRMs.}
  \label{receivers}
\end{figure*}

The Passive Receiver (PR) concept is introduced to simplify the complex receiver structures \cite{pierobon2011diffusion}. The PR is often approximated as a sphere that efficiently absorbs molecules into the IDRM's interior. In this simplification, the receiver does not actively participate in the propagation of molecules but effectively directs them into the IDRM. This concept aligns well with the absorption requirements of the DAC and serves as a suitable model for its efficient absorption by the IDRM. The IDRM units' components are visually represented in Fig. \ref{receivers}.

Another type of receiver concept that can be considered is the Absorbing Receiver. This receiver approximated as a sphere, selectively absorbs the desired molecule upon interacting within the corresponding space. In this receiver type, the desired molecule is solely sensed by the receiver but does not enter the MCS. This receiver design is particularly relevant for the dopamine detection aspect within our IDRM.  We consider dopamine release as a  probability distribution similar to a Poisson distribution that can partially clarify the functioning of dopamine release in these receivers. In both types of receivers, it is assumed that they are densely located on the surface of the capsule, exhibiting an infinitely high absorption rate. The behavior of molecules follows the principles of 3D free diffusion within the uniform flow. The number of molecules adsorbed within the spherical space of the PR can be regarded as a nonstationary Poisson process.

\begin{equation}\label{equ1}
   N_{Rx\mid PA(t)}\sim Poission(\lambda_{Rx}(t)).
\end{equation}

The parameter $\lambda$ in the Poisson process exhibits temporal variation, and its expression can be represented as:

\begin{equation}\label{equ2}
 \lambda_{Rx}(t)= \lambda_{noise}+\sum_{j=1}^{\frac{t}{T_s}+1} P_{obs}(t-(j-i)T_s),    
\end{equation}
Where $T_s$ represents the duration of the sampling time, and $\lambda_{noise}$ corresponds to the additive static noise that partially reflects the influence of the surrounding environment on the side Rx.

The function $P_{obs}$ quantifies the number of molecules reaching the Rx sampling space within the time interval from t=0 to t is expressed as

\begin{equation}\label{equ3}
 P_{obs}=\frac{V_{Rx}}{{(4\pi Dt)}^{1.5}}([C_{DAC}]),   
\end{equation}
where $[C_{DAC}]$ represents the concentration of DAC near the capsule, a crucial parameter for assessing the IDRM's DAC molecule availability and abundance. $V_{Rx}=\frac{4}{3} \pi d^3_{Rx}$ is the volume of the spherical receiver. $d_{Rx}$ is the radius of the receiver, which characterizes its capacity and the spatial extent to which it can absorb molecules. $D$ is the diffusion coefficient, determining the rate at which molecules disperse within the surrounding medium.


\section{Delivering Dopamine Compound to the storage part of Intelligent rate
Dopamine Modulator}

\begin{figure*}[t]
  \centering
  \includegraphics[width=\textwidth]{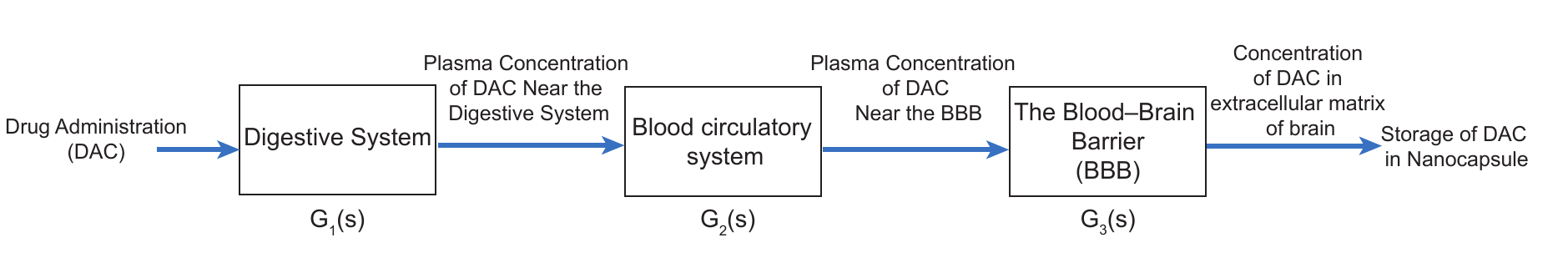}
  \caption{DDS components for transferring DAC to IDRMs.}
  \label{DDS2}
\end{figure*}

As previously indicated, DAC would be employed to cross the IDRM and undergo transformation into dopamine within its confines. Practical mechanisms used by Drug Delivery Systems (DDS) will be implemented to transport the DAC to the target-absorbing receptor located within the IDRM. Exogenous dopamine enters the body via oral drug administration. In this section, the drug transfer from ingestion to reaching the capsule is modeled. As with all orally administered medications, DAC is ingested in pill form and metabolized by the gastrointestinal tract. Within the gastrointestinal tract, absorption occurs as the villi take up the substance, facilitating its diffusion into the bloodstream. Afterward, as it progresses through the blood circulatory system, it eventually reaches BBB. It has been previously noted that the structure of DAC should be designed to facilitate crossing BBB. Subsequently, it diffuses into the extracellular fluid within the brain to access the IDRM situated in the synaptic cleft, connecting the SNc and the striatum. Incorporating DAC into the IDRM occurs via the absorbing receptor, leading to its subsequent conversion into dopamine.
Furthermore, this section aims to delineate components of DDS for delivering DAC to the IDRM, and their mathematical relationships will be partially modeled. While this segment aligns with MC system principles, it cannot be considered an MC system due to its non-information-transmitting nature. However, its sole function is to facilitate the charging of the IDRM with dopamine. A block diagram of the components of DDS is indicated in Fig.~\ref{DDS2}. In examining these pivotal elements, valuable insights into DDS and its application in neurodegenerative diseases, such as Parkinson’s disease, are aimed to be provided, explicitly emphasizing the potential advantages of localized dopamine release in the SNc region.

\subsubsection{DAC concentration in plasma level}
Since the precise structure of DAC has not yet been determined, the behavioral characteristics of the most closely related pharmaceutical, Levodopa, have been employed to model this particular aspect. Experimental measurements of blood plasma levels for Levodopa have been reported in several scholarly journals \cite{deleu2002clinical, hacisalihzade1989optimization}.  A systemic approach has been taken to characterize the behavior of the Levodopa in this part. Drug administration is modeled with its dosage being used as the input and the drug concentration in the blood as the output. The input associated with each drug administration is represented as a unit impulse signal with an amplitude constrained by the drug dosage. By observing the distribution of Levodopa, it is evident by observing experimental data that this behavior resembles the impulse response of a second-order transfer function system with a delay, as 

\begin{equation}\label{equ4}
    G_1(s)=\frac{ke^{-T_0s}}{(1+sT_1)(1+sT_2)}.
\end{equation}
where the amplification coefficient, denoted as $k$, is expressed in milligrams (mg), and the time constants $T_1$ and $T_2$ are quantified in hours. A delay indicates the time it takes for Levodopa to be absorbed into the bloodstream, and $T_0$ represents a delay indicating the duration required to absorb Levodopa into the bloodstream.

\subsubsection{DAC concentration in blood circulation system}

In order to comprehensively capture the behavior of the drug within the blood circulatory system, it is essential to consider the intricate processes involved in its absorption and distribution. Following absorption in the vessels near the intestine, the drug traverses through the blood circulatory system , reaching various body parts before ultimately reaching the brain. The measured blood plasma level in the preceding section was related to an area near the digestive tract. It is evident that DAC needs to traverse from that area to the BBB, where a portion of it will ultimately reach. Therefore, it is possible to represent this particular segment as a delayed system with an attenuation factor. The utilization of this modeling methodology is explained by the time delay in which the drug-containing blood reaches the BBB from the digestive system. Additionally, a fraction of the drug's concentration in the blood will concurrently reach the vicinity of the BBB, facilitating its translocation, which can be obtained as

\begin{equation}\label{equ5}
   G_2(s)=ae^{-sT_3},   
\end{equation}
where $a$ follows the condition $0 < a < 1$, represents the attenuation coefficient of DAC as it acrosses from the vicinity of the intestine to the BBB, while $T_3$ indicates the time delay of DAC within the blood circulatory system.

\subsubsection{DAC concentration across the Blood-Brain Barrier}

The blood-brain barrier serves as a selective barrier, facilitating the translocation of specific substances while impeding the ingress of others. The administered drug should contain a dopaminergic compound capable of crossing this barrier. Upon successful transcytosis, the drug exhibits an elevation in concentration across the entirety of cerebral regions. The pharmaceutical levodopa operates on analogous principles, constituting its foremost limitation. Notably, it elicits a pervasive elevation in dopamine concentrations across all cerebral domains. While it does manifest advantageous effects within the SNc region in alleviating symptoms of PD, it concurrently perturbs regions housing undisturbed dopaminergic neuronal populations. This multifaceted approach to pharmacological intervention engenders a spectrum of collateral effects, entangling neurons in a drug-dependent state and diminishing endogenous dopamine excretion with the gradual emergence of neuronal resilience. Hence, the management of PD in its advanced stages necessitates dose escalation, a therapeutic approach often rendered impractical in clinical settings due to the drug's pronounced side effects upon administration of a pharmaceutical compound comprising dopamine. Its neurotransmitter effect cannot be manifest in other cerebral regions. The absorption of this compound in the targeted IDRMs occurs exclusively within the SNc region.
Furthermore, it is within the transducer compartment contained within these IDRMs where the conversion to dopamine occurs. Therefore, utilizing IDRMs allows the targeted dopamine release exclusively within the SNc region. 

The DAC effectively crosses the BBB within this particular segment, thereby attaining entry into the cerebral region. Several articles have put different hypotheses regarding how the BBB is traversed \cite{shamloo2016computing}. However, it seems that this behavior can also be effectively modeled using a first-order transfer function.
The BBB response to DAC through the blood-brain barrier can be compared to the behavior of first-order capacitor charging circuits.
Therefore, it is reasonable to explore utilizing a transfer function to depict this transition accurately. It is defined as

\begin{equation}\label{equ6}
    G_3(s)=\frac{\beta }{s+\beta},
\end{equation}
Where $\frac{1}{\beta}$  represents the time-constant value of the DAC passage across the BBB. Since the precise type and structure of the drug have not yet been accurately determined, the performance of the BBB in allowing the passage of DAC with different structures may vary.

\subsubsection{DAC propagation in the extracellular matrix of the brain}

In the preceding phase, the penetration of DAC from the BBB into the cerebral tissue was shown. At this stage, the distribution of DAC within the brain's extracellular matrix. The brain's extracellular matrix is an intercellular space where various chemical compounds diffuse, spreading physiological information across cells (neurons and non-neurons). Fraction and tortuosity are the two parameters that affect the diffusion process \cite{veletic2019molecular}. The volume fraction  is the ratio of total tissue to the extracellular matrix and can be determined as

\begin{equation}\label{equ7}
    \alpha={\mathcal{V}_{Extracellular\hspace{1mm} Matrix}}/{\mathcal{V}_{Tissue}}
\end{equation}
where $\mathcal{V}_{Extracellular\hspace{1mm} Matrix}$ represents the volume of the extracellular  matrix, while $\mathcal{V}_{Tissue}$ signifies the total tissue volume within a specific brain area. The typical value range for $\alpha$ \hspace{2mm} is \hspace{2mm} $0.1\leq\alpha\leq 0.3 $. Tortuosity in biological contexts refers to the total impedance encountered by a complex medium compared to an unobstructed environment and is defined as

\begin{equation}\label{equ8}
    \lambda=\sqrt{D/D^{\ast}}.
\end{equation}

This impedance decreases the effective diffusion of DAC compared to the unconstrained diffusion coefficient indicated as D. The effective diffusivity of DAC within the brain's extracellular matrix, denoted as $D^{\ast}$ is influenced by the tortuosity, approximately constant at $\lambda\simeq1.6$.
$c_{DAC}(t,x)$ represents the concentration of DAC in the extracellular matrix dependent on both time (t) and spatial coordinates ($x$). The diffusive flow is described as \cite{veletic2019molecular},

\begin{align}\label{equ9}
   \frac{\partial c{_{DAC}(t,x)}}{\partial t} = \underbrace{\frac{D}{\lambda^2} c{_{DAC}(t,x)+\frac{c{_{DAC}(t_0,x_0)}}{\alpha}}}_\text{diffusion}- \\ \nonumber \underbrace{\frac{\mathnormal{f(c_{DAC})}}{\alpha}}_\text{uptake}
   - \underbrace{\mathnormal{v} \nabla c{_{DAC}(t,x)}}_\text{bulk flow}
\end{align}

Through calculation explained in reference \cite{sykova2008diffusion} (\ref{equ9}) is simplifies as:

\begin{equation}\label{equ10}
  c{_{DAC}(t,x)}=\frac{\mathnormal{Q}\lambda^2}{4\pi \mathnormal{D}\alpha r} \mathrm{erfc} \left( \frac{r\lambda}{2\sqrt{\mathnormal{D}|t|}}\right).
\end{equation}

In the context of one dimension, $r=x$, whereas in the three-dimensional but in the context of three dimensions, represents as $r=\sqrt{x^2+y^2+z^2}$. The analysis of changes in concentration over time will be conducted for variable r. Considering the IDRMs' fixed distance from the BBB to the striatum, r is a numerical value to compute concentrations across all time intervals. Here, $c_{DAC}(t_0,0)=\mathnormal{Q}$ represents the amount of DAC dust from the BBB immediately after the BBB in the brain's extracellular matrix.
It appears that the release of different DAC structures would be adequately demonstrative with the parameters in \eqref{equ10}. However, a crucial aspect to consider here is the distance from the BBB to the IDRM site. Given that the precise determination of this distance is not feasible and the positioning of the IDRM from the BBB could vary, the average of this distance is considered. Deriving this average requires a separate study.

\section{Numerical results}

\begin{figure}[t]
  \centering
  \includegraphics[width=\columnwidth]{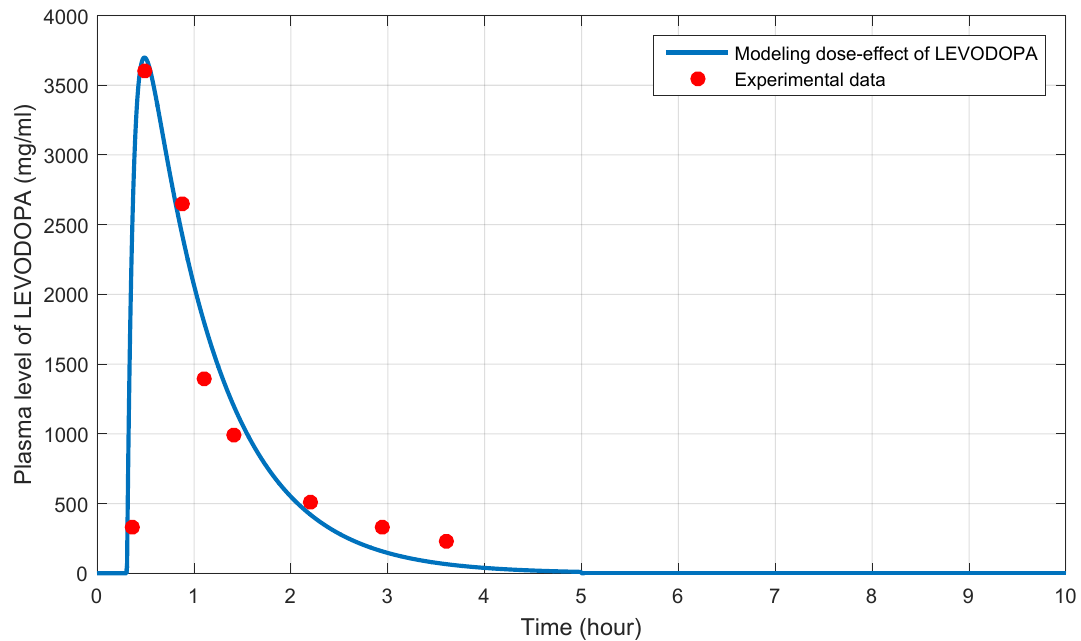}
  \caption{Blood plasma level of Levodopa during the time\cite{hacisalihzade1989optimization}.}\label{levodopa}
\end{figure}

In this section, numerical simulations are performed using the analytical results to evaluate introduced DDS behavior, which is specified from drug (DAC) ingestion to reaching the IDRM. The main purpose of DDS is charging the IDRM with needed dopamine. The proposed IDRM is determined with system parameters shown in table \ref{table1}. In the first
stage, the level of Levodopa in the blood has been measured multiple times after administering 125 mg of Levodopa capsules to identify the parameters of $G_1(s)$ (see  (\ref{equ4})). The parameter values have been determined using the least square method \cite{haeri2005modeling}, and the system’s response to a single drug dose alongside the experimentally obtained data is plotted in Fig.~\ref{levodopa}. Parameters are determined for the DDS using the least square method as $k=1418$ mg, $T_1=0.0547$ hour, $T_2=0.6073$ hour, and $T_0=0.2461$ hour. This initial phase of the DDS  entails the administration of the pharmaceutical compound and subsequent tracking of its impact in the bloodstream. The parameter values of the levodopa were utilized for this phase since the structure of DAC has yet to be determined, and it is perceived that the behavior of DAC might  have some similarity to levodopa's behavior.

\begin{figure}[t]
  \begin{center}
  \includegraphics[width=\columnwidth]{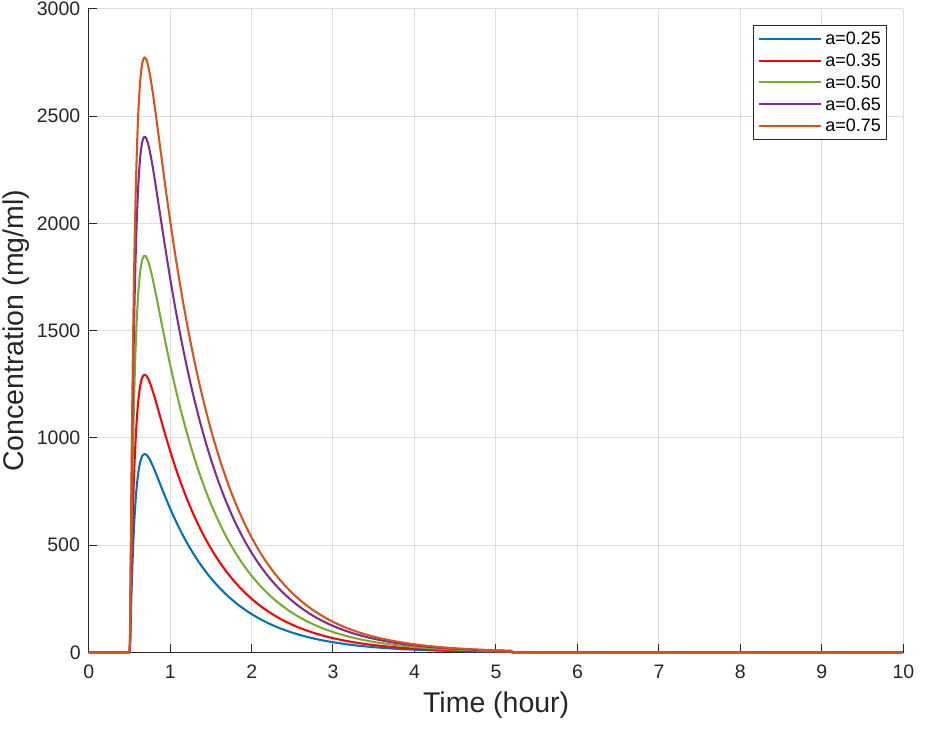}
  \caption{Pharmacokinetics of DAC across the circulatory system due to varying value. }\label{result}
  \end{center}
\end{figure}

\begin{table*}[ht]  
    \centering  
    \footnotesize  
    \caption{List of parameters used in the analysis.}
    \label{table1}  
    \begin{tabular}{|l|c|c|}  
        \hline
        Parameter & Symbol & Value \\
        \hline
        Amplification coefficient & k & 1418 mg \\
        Time constants in $G_1(s)$ & $T_1$, $T_2$ & 0.0547, 0.6073 hour \\
        Time delay in $G_1(s)$ & $T_0$ & 0.2461 hour \\
        Time constant in $G_2(s)$ & $T_3$ & 0.2 hour \\
        Attenuation coefficient in $G_2(s)$ & a & 0.25, 0.35, 0.50, 0.60, 0.75 \\
        Time constant in $G_3(s)$ & $\beta$ & 0.50, 0.75, 1, 1.25, 1.50 hour\(^{-1}\) \\
        Volume fraction & $\alpha$ & 0.2 \\
        Diffusion coefficient & $D$ & 15 $\mu m^2/s$ \\
        Distance from the BBB to IDRM & r & 1, 1.20, 1.30, 1.40, 1.50 mm \\
        Tortuosity & $\lambda$ & 1.6 \\
        \hline
    \end{tabular}
\end{table*}

parameters of $G_2(s)$ is considered in (5) as [0.25, 035, 0.50, 0.60, 0.75] and T3=0.2 hour. To analyze the behavior of a drug in the blood circulation system. It is important to note that the duration of this time delay can vary among individuals; however, based on average observations, a time delay of approximately 0.2 hour is deemed appropriate. The result of the blood circulation system is shown in Fig. \ref{result}.



\begin{figure}[t]
  \begin{center}
  \includegraphics[width=\columnwidth]{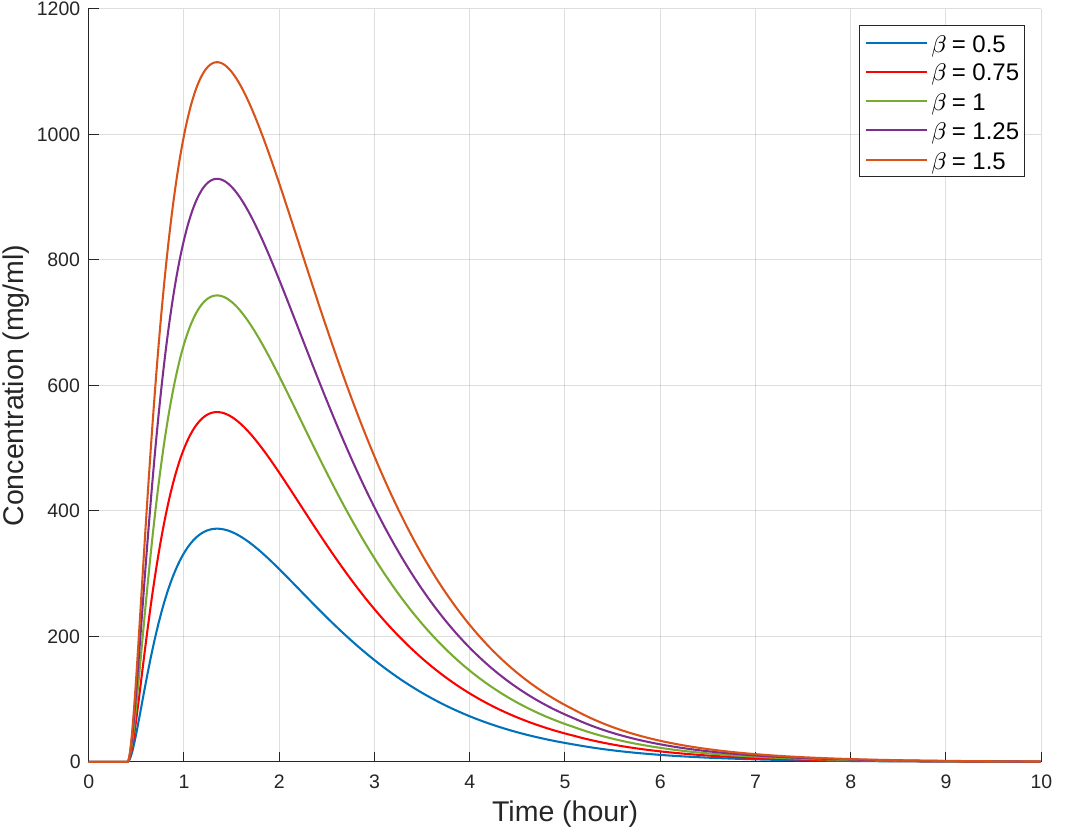}
  \caption{Pharmacokinetics of DAC across BBB due to changing  time-constant values.}\label{beta}
  \end{center}
\end{figure}

It is important to note that the structure of the DAC is unspecified; the time constant of crossing the BBB cannot be accurately predicted. To simulate the nature of the flow behavior of the DAC, $\beta$ was simulated for several different values. The simulation results for $\beta$ are varied as [0.5, 0.75, 1, 1.25, 1.50] illustrated in Fig. \ref{beta}. If the structure of the drug becomes known later in the studies, it will be possible to calculate this value precisely.


\begin{figure}[t]
  \begin{center}
  \includegraphics[width=\columnwidth]{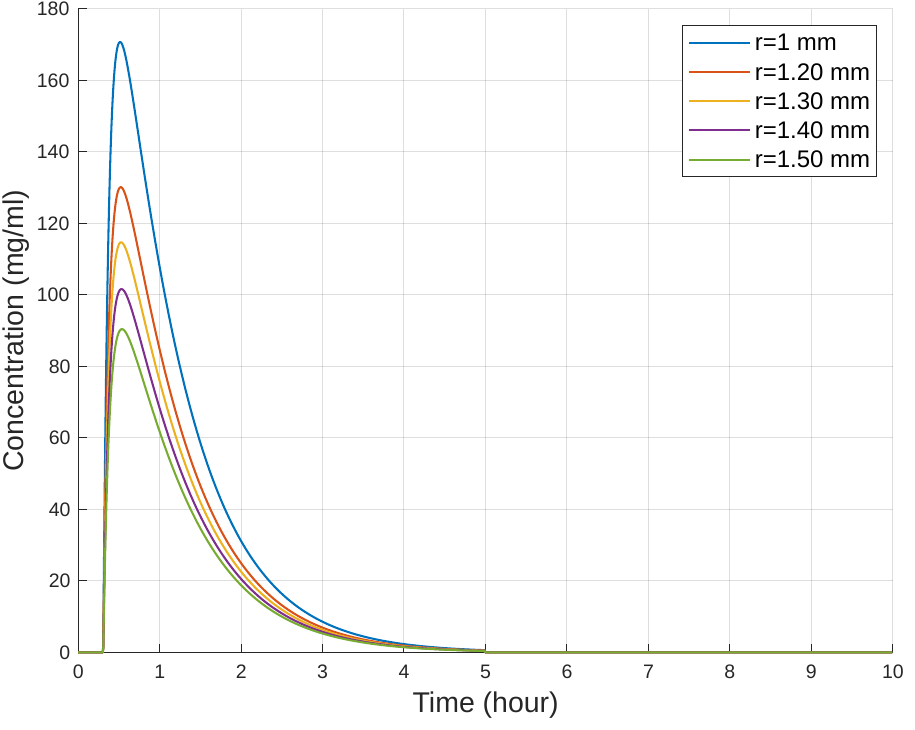}
  \caption{ Concentration of DAC delivery to IDRM within the brain's extracellular matrix across varied distances from the BBB.}\label{brain}
  \end{center}
\end{figure}

To measure the concentration of the DAC quantitatively in the extracellular matrix of the brain, the location of the IDRM in various $r$ has been considered and calculated. Therefore, the one-dimensional distance r is alternated as [1mm, 1.20mm, 1.30mm, 1.40mm, 1.50mm] depicted in Fig. \ref{brain}.


\begin{figure}[htbp]
  \begin{center}
  \includegraphics[width=\columnwidth]{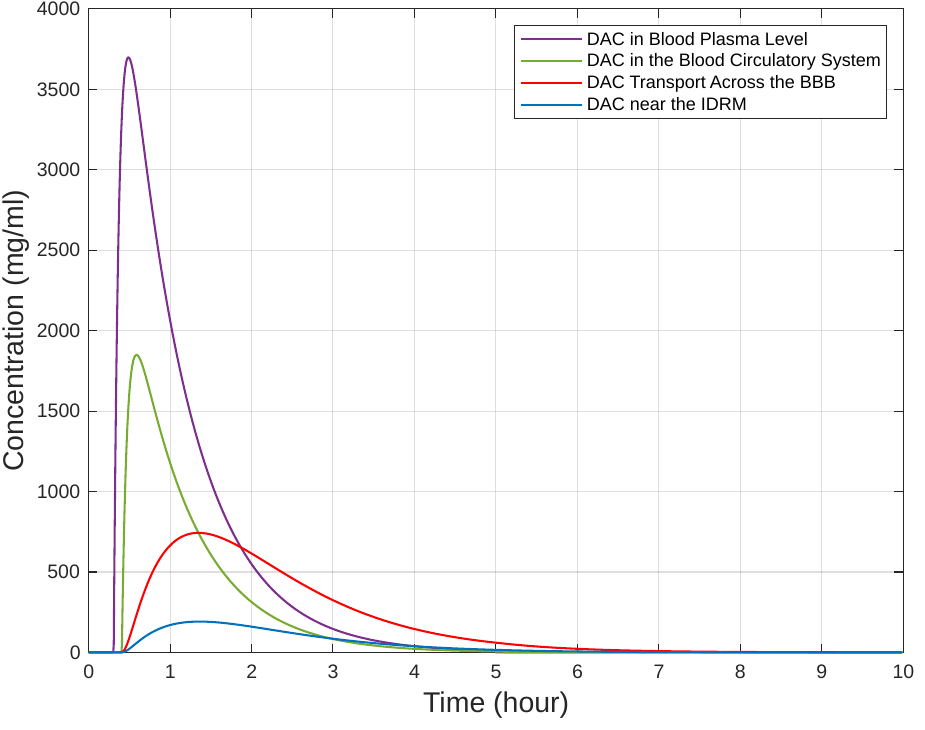}
  \caption{Pharmacokinetics of DAC Across DDS components with considering average values.}\label{DDS}
  \end{center}
\end{figure}

In Fig. \ref{DDS}, a decrease in DAC concentration across various regions of the DDS is depicted, showing the declining concentration as it is delivered through different regions. In every scenario, average parameters have been considered.  The final concentration is the concentration of DAC in the vicinity of the IDRMs over time. This suggests that regular intake of DAC, which is received via IDRMs, can effectively ensure that the required dopamine is available in the BG.

In this study, we proposed the treatment method by focusing on the main cause of PD (the death of SNc neurons and the lack of dopamine secretion). For this purpose, a network of small units based on MC known as IDRMs are designed to replace the dead SNc neurons and placed between the SNc and the striatum. One natural indicator of SNc activity is the release of dopamine. However, due to the death of specific neurons in the SNc, the level of naturally occurring dopamine release may be insufficient. The presented IDRM system can compensate for this deficiency by secreting dopamine in such cases. The mechanism of these IDRMs will be such that the dopamine stored inside the IDRM will be released when the receptor of the IDRM detects the presence of dopamine in the area between SNc and the striatum. Thus, the required dopamine will be released when needed. This approach will effectively compensate for the dopamine shortage brought on by the demise of SNc cells. The symptoms of the disease will be well suppressed, and the physiological function of the SNc region will be preserved. The primary issue in this method is how to store dopamine in the IDRM. The patient will be prescribed medicine containing dopamine compounds to deal with this issue. The DAC enters the bloodstream after passing through the digestive system. Next, it enters the brain after passing through the BBB. The transducer in the IDRM should transform DAC into dopamine while it is consumed. This dopamine is stored in the vesicles in the IDRM and can be released when needed. The presented method minimizes the side effects of the drug because dopamine will not be released in other areas of the brain. It can be confidently stated that this treatment approach aims to target the root cause of the disease and is not merely aimed at reducing symptoms, unlike other treatments. Our simulations and analyses showed that by using DDS, we can be confident that the needed amount of dopamine will be stored in IDRM by using DAC tablets at the proper times.

In this paper, we proposed a novel approach to treatment for Parkinson's disease (PD) targeting the main cause. By placing IDRMs for MC-based systems to release dopamine, our approach targets the degeneration-induced dopamine deficiency caused by the loss of SNc neurons. These IDRMs release dopamine to restore decreased dopamine levels caused by the degeneration of SNc neurons. Our scenario involves utilizing existing biological mechanisms that still naturally transport substances in the body properly. They are directly placed between the SNc and the striatum. They are capable of detecting dopamine levels and releasing stored dopamine when necessary. This mechanism efficiently decreases disease symptoms while maintaining the physiological functioning of the SNc region. This approach directly targets the fundamental cause of the disease by restoring dopamine transmission in the basal ganglia.
Moreover, the natural communication mechanism ensures compatibility with the body's physiological structure and holds promise for advanced treatments in PD. It addresses the limitations of current treatments and underscores the importance of tackling the underlying cause of PD. Further research and clinical investigations are essential to validate the efficacy and safety of this proposed treatment approach. Through continued exploration of alternative and complementary therapies, the prospect of improved disease management and enhanced quality of life for patients becomes increasingly promising.


%





\ifCLASSOPTIONcaptionsoff
  \newpage
\fi





\bibliographystyle{IEEEtran}
\bibliography{Bibliography}

\begin{thebibliography}{10}
\providecommand{\url}[1]{#1}
\csname url@rmstyle\endcsname
\providecommand{\newblock}{\relax}
\providecommand{\bibinfo}[2]{#2}
\providecommand\BIBentrySTDinterwordspacing{\spaceskip=0pt\relax}
\providecommand\BIBentryALTinterwordstretchfactor{4}
\providecommand\BIBentryALTinterwordspacing{\spaceskip=\fontdimen2\font plus
\BIBentryALTinterwordstretchfactor\fontdimen3\font minus \fontdimen4\font\relax}
\providecommand\BIBforeignlanguage[2]{{%
\expandafter\ifx\csname l@#1\endcsname\relax
\typeout{** WARNING: IEEEtran.bst: No hyphenation pattern has been}%
\typeout{** loaded for the language `#1'. Using the pattern for}%
\typeout{** the default language instead.}%
\else
\language=\csname l@#1\endcsname
\fi
#2}}

\bibitem{olanow2009scientific}
C.~W. Olanow, M.~B. Stern, and K.~Sethi, ``The scientific and clinical basis for the treatment of parkinson disease (2009),'' \emph{Neurology}, vol.~72, no. 21 Supplement 4, pp. S1--S136, 2009.

\bibitem{przedborski2017two}
S.~Przedborski, ``The two-century journey of parkinson disease research,'' \emph{Nature Reviews Neuroscience}, vol.~18, no.~4, pp. 251--259, 2017.

\bibitem{chaudhuri2009non}
K.~R. Chaudhuri and A.~H. Schapira, ``Non-motor symptoms of parkinson's disease: dopaminergic pathophysiology and treatment,'' \emph{The Lancet Neurology}, vol.~8, no.~5, pp. 464--474, 2009.

\bibitem{bohnen2022discussion}
N.~I. Bohnen, R.~M. Costa, W.~T. Dauer, S.~A. Factor, N.~Giladi, M.~Hallett, S.~J. Lewis, A.~Nieuwboer, J.~G. Nutt, K.~Takakusaki, \emph{et~al.}, ``Discussion of research priorities for gait disorders in parkinson's disease,'' \emph{Movement disorders}, vol.~37, no.~2, pp. 253--263, 2022.

\bibitem{aarsland2021parkinson}
D.~Aarsland, L.~Batzu, G.~M. Halliday, G.~J. Geurtsen, C.~Ballard, K.~Ray~Chaudhuri, and D.~Weintraub, ``Parkinson disease-associated cognitive impairment,'' \emph{Nature Reviews Disease Primers}, vol.~7, no.~1, p.~47, 2021.

\bibitem{tran2018levodopa}
T.~N. Tran, T.~N. Vo, K.~Frei, and D.~D. Truong, ``Levodopa-induced dyskinesia: clinical features, incidence, and risk factors,'' \emph{Journal of Neural Transmission}, vol. 125, pp. 1109--1117, 2018.

\bibitem{dale2022evoked}
J.~Dale, S.~L. Schmidt, K.~Mitchell, D.~A. Turner, and W.~M. Grill, ``Evoked potentials generated by deep brain stimulation for parkinson's disease,'' \emph{Brain stimulation}, 2022.

\bibitem{haeri2005modeling}
M.~Haeri, Y.~Sarbaz, and S.~Gharibzadeh, ``Modeling the parkinson's tremor and its treatments,'' \emph{Journal of theoretical biology}, vol. 236, no.~3, pp. 311--322, 2005.

\bibitem{armstrong2020diagnosis}
M.~J. Armstrong and M.~S. Okun, ``Diagnosis and treatment of parkinson disease: a review,'' \emph{Jama}, vol. 323, no.~6, pp. 548--560, 2020.

\bibitem{little2021smartphones}
M.~A. Little, ``Smartphones for remote symptom monitoring of parkinson’s disease,'' \emph{Journal of Parkinson's Disease}, vol.~11, no.~s1, pp. S49--S53, 2021.

\bibitem{di2018quantitative}
L.~Di~Biase, S.~Summa, J.~Tosi, F.~Taffoni, M.~Marano, A.~Cascio~Rizzo, F.~Vecchio, D.~Formica, V.~Di~Lazzaro, G.~Di~Pino, \emph{et~al.}, ``Quantitative analysis of bradykinesia and rigidity in parkinson’s disease,'' \emph{Frontiers in neurology}, vol.~9, p. 121, 2018.

\bibitem{bloem2015nonpharmacological}
B.~R. Bloem, N.~M. de~Vries, and G.~Ebersbach, ``Nonpharmacological treatments for patients with parkinson's disease,'' \emph{Movement Disorders}, vol.~30, no.~11, pp. 1504--1520, 2015.

\bibitem{wang2022efficacy}
Y.~Wang, X.~Sun, F.~Li, Q.~Li, and Y.~Jin, ``Efficacy of non-pharmacological interventions for depression in individuals with parkinson's disease: A systematic review and network meta-analysis,'' \emph{Frontiers in Aging Neuroscience}, vol.~14, p. 1050715, 2022.

\bibitem{modi2009nanotechnological}
G.~Modi, V.~Pillay, Y.~E. Choonara, V.~M. Ndesendo, L.~C. du~Toit, and D.~Naidoo, ``Nanotechnological applications for the treatment of neurodegenerative disorders,'' \emph{Progress in Neurobiology}, vol.~88, no.~4, pp. 272--285, 2009.

\bibitem{spuch2012advances}
C.~Spuch, O.~Saida, and C.~Navarro, ``Advances in the treatment of neurodegenerative disorders employing nanoparticles,'' \emph{Recent patents on drug delivery \& formulation}, vol.~6, no.~1, pp. 2--18, 2012.

\bibitem{biletic2020nanonetworks}
M.~Biletic, F.~H. Juwono, and L.~Gopal, ``Nanonetworks and molecular communications for biomedical applications,'' \emph{IEEE Potentials}, vol.~39, no.~3, pp. 25--30, 2020.

\bibitem{akan2016fundamentals}
O.~B. Akan, H.~Ramezani, T.~Khan, N.~A. Abbasi, and M.~Kuscu, ``Fundamentals of molecular information and communication science,'' \emph{Proceedings of the IEEE}, vol. 105, no.~2, pp. 306--318, 2016.

\bibitem{dressler2010bio}
F.~Dressler and O.~B. Akan, ``Bio-inspired networking: from theory to practice,'' \emph{IEEE Communications Magazine}, vol.~48, no.~11, pp. 176--183, 2010.

\bibitem{atakan2012body}
B.~Atakan, O.~B. Akan, and S.~Balasubramaniam, ``Body area nanonetworks with molecular communications in nanomedicine,'' \emph{IEEE Communications Magazine}, vol.~50, no.~1, pp. 28--34, 2012.

\bibitem{atakan2012bio}
B.~Atakan and O.~B. Akan, ``Bio-inspired cross-layer communication and coordination in sensor and vehicular actor networks,'' \emph{IEEE transactions on vehicular technology}, vol.~61, no.~5, pp. 2185--2193, 2012.

\bibitem{vegh2022towards}
J.~V{\'e}gh and {\'A}.~J. Berki, ``Towards generalizing the information theory for neural communication,'' \emph{Entropy}, vol.~24, no.~8, p. 1086, 2022.

\bibitem{veletic2019molecular}
M.~Veleti{\'c}, M.~T. Barros, I.~Balasingham, and S.~Balasubramaniam, ``A molecular communication model of exosome-mediated brain drug delivery,'' in \emph{Proceedings of the Sixth Annual ACM International Conference on Nanoscale Computing and Communication}, 2019, pp. 1--7.

\bibitem{akan2021information}
O.~B. Akan, H.~Ramezani, M.~Civas, O.~Cetinkaya, B.~A. Bilgin, and N.~A. Abbasi, ``Information and communication theoretical understanding and treatment of spinal cord injuries: State-of-the-art and research challenges,'' \emph{IEEE Reviews in Biomedical Engineering}, vol.~16, pp. 332--347, 2021.

\bibitem{dressler2010survey}
F.~Dressler and O.~B. Akan, ``A survey on bio-inspired networking,'' \emph{Computer networks}, vol.~54, no.~6, pp. 881--900, 2010.

\bibitem{malak2012molecular}
D.~Malak and O.~B. Akan, ``Molecular communication nanonetworks inside human body,'' \emph{Nano Communication Networks}, vol.~3, no.~1, pp. 19--35, 2012.

\bibitem{malak2013communication}
------, ``A communication theoretical analysis of synaptic multiple-access channel in hippocampal-cortical neurons,'' \emph{IEEE Transactions on communications}, vol.~61, no.~6, pp. 2457--2467, 2013.

\bibitem{ramezani2017communication}
H.~Ramezani and O.~B. Akan, ``A communication theoretical modeling of axonal propagation in hippocampal pyramidal neurons,'' \emph{IEEE transactions on nanobioscience}, vol.~16, no.~4, pp. 248--256, 2017.

\bibitem{ramezani2018impacts}
------, ``Impacts of spike shape variations on synaptic communication,'' \emph{IEEE transactions on nanobioscience}, vol.~17, no.~3, pp. 260--271, 2018.

\bibitem{ramezani2017importance}
------, ``Importance of vesicle release stochasticity in neuro-spike communication,'' in \emph{2017 39th Annual International Conference of the IEEE Engineering in Medicine and Biology Society (EMBC)}.\hskip 1em plus 0.5em minus 0.4em\relax IEEE, 2017, pp. 3343--3347.

\bibitem{khan2019impact}
T.~Khan, H.~Ramezani, N.~A. Abbasi, and O.~B. Akan, ``Impact of long term plasticity on information transmission over neuronal networks,'' \emph{IEEE transactions on nanobioscience}, vol.~19, no.~1, pp. 25--34, 2019.

\bibitem{lang2004challenges}
A.~E. Lang and J.~A. Obeso, ``Challenges in parkinson's disease: restoration of the nigrostriatal dopamine system is not enough,'' \emph{The Lancet Neurology}, vol.~3, no.~5, pp. 309--316, 2004.

\bibitem{xie2019nanomaterial}
J.~Xie, Z.~Shen, Y.~Anraku, K.~Kataoka, and X.~Chen, ``Nanomaterial-based blood-brain-barrier (bbb) crossing strategies,'' \emph{Biomaterials}, vol. 224, p. 119491, 2019.

\bibitem{lu2023artificial}
J.~Lu, G.~Jiang, H.~Zhang, B.~Qian, H.~Zhu, Q.~Gu, Y.~Yan, J.~Z. Liu, B.~D. Freeman, L.~Jiang, \emph{et~al.}, ``An artificial sodium-selective subnanochannel,'' \emph{Science Advances}, vol.~9, no.~4, p. eabq1369, 2023.

\bibitem{alabrahim2022polymeric}
O.~A.~A. Alabrahim and H.~M. E.-S. Azzazy, ``Polymeric nanoparticles for dopamine and levodopa replacement in parkinson's disease,'' \emph{Nanoscale advances}, vol.~4, no.~24, pp. 5233--5244, 2022.

\bibitem{kilinc2013receiver}
D.~Kilinc and O.~B. Akan, ``Receiver design for molecular communication,'' \emph{IEEE Journal on Selected Areas in Communications}, vol.~31, no.~12, pp. 705--714, 2013.

\bibitem{kuscu2019transmitter}
M.~Kuscu, E.~Dinc, B.~A. Bilgin, H.~Ramezani, and O.~B. Akan, ``Transmitter and receiver architectures for molecular communications: A survey on physical design with modulation, coding, and detection techniques,'' \emph{Proceedings of the IEEE}, vol. 107, no.~7, pp. 1302--1341, 2019.

\bibitem{pierobon2011diffusion}
M.~Pierobon and I.~F. Akyildiz, ``Diffusion-based noise analysis for molecular communication in nanonetworks,'' \emph{IEEE Transactions on signal processing}, vol.~59, no.~6, pp. 2532--2547, 2011.

\bibitem{deleu2002clinical}
D.~Deleu, M.~G. Northway, and Y.~Hanssens, ``Clinical pharmacokinetic and pharmacodynamic properties of drugs used in the treatment of parkinson’s disease,'' \emph{Clinical pharmacokinetics}, vol.~41, pp. 261--309, 2002.

\bibitem{hacisalihzade1989optimization}
S.~Hacisalihzade, M.~Mansour, and C.~Albani, ``Optimization of symptomatic therapy in parkinson's disease,'' \emph{IEEE transactions on biomedical engineering}, vol.~36, no.~3, pp. 363--372, 1989.

\bibitem{shamloo2016computing}
A.~Shamloo, M.~Z. Pedram, H.~Heidari, and A.~Alasty, ``Computing the blood brain barrier (bbb) diffusion coefficient: A molecular dynamics approach,'' \emph{Journal of Magnetism and Magnetic Materials}, vol. 410, pp. 187--197, 2016.

\bibitem{sykova2008diffusion}
E.~Sykov{\'a} and C.~Nicholson, ``Diffusion in brain extracellular space,'' \emph{Physiological reviews}, vol.~88, no.~4, pp. 1277--1340, 2008.

\end{thebibliography}

\vfill


\end{document}